\documentclass[
preprint,%
english,%
aps,%
prl,%
showpacs,%
superscriptaddress,%
floats,%
amssymb]{revtex4}
\usepackage[T1]{fontenc}
\usepackage[latin1]{inputenc}

\makeatletter

\usepackage{graphicx}
\usepackage{amsmath}
\usepackage{amssymb}
\bibstyle{apsrev.bib}

\newcommand{\beq}{\begin{equation}}
\newcommand{\eeq}{\end{equation}}

\newcommand{\comment}[1]{}

\usepackage{babel}
\makeatother
\begin{document}

\title{Spin Glass Stiffness in zero Dimensions}

\author{Stefan Boettcher\footnote{http://www.physics.emory.edu/faculty/boettcher} }

\affiliation{Physics Department, Emory University, Atlanta, Georgia 30322, USA}

\begin{abstract}
A unique analytical result for the Migdal-Kadanoff hierarchical
lattice is obtained. The scaling of the defect energy for a
zero-dimensional spin glass is derived for a bond distribution that is
continuous at the origin. The value of the {}``stiffness'' exponent in
zero dimensions, $y_{0}=-1,$ corresponds to the value also found in
one dimension. This result complements and completes earlier findings
for $y_{d}$ at $d>0.$
\end{abstract}


\maketitle
A quantity of fundamental importance for the modeling of amorphous
magnetic materials through spin glasses~\cite{F+H} is the
{}``stiffness'' exponent $y$~\cite{Southern77,Bray84}. The stiffness
of a spin configuration describes the typical rise in magnetic energy
$\Delta E$ due to an induced defect-interface of size $L$. In a glassy
system, the potential energy function resembles a high-dimensional
mountain landscape over its variables~\cite{Frauenfelder96}. Any
defect-induced displacement of size $L$ in such a landscape may move a
system equally likely up or down in energy $\Delta E$. Averaging over
many incarnations of such a system then results in a typical energy
scale
\begin{eqnarray} <|\Delta E|>\sim L^{y}\quad(L\rightarrow\infty).
\label{yeq}
\end{eqnarray} 
The importance of this exponent for small excitations
in disordered spin systems has been discussed in many
contexts~\cite{Southern77,Bray84,Fisher86,Krzakala00,Palassini00,Bouchaud03,F+H}.
In particular, it signifies a renormalized coupling strength (across
any hypothetical interface) between regions in space separated by a
distance $L$~\cite{Bray84}: If $y_{d}>0,$ regions in space are
strongly coupled at low temperature and spin glass ordering ensues,
i. e. $T_{g}>0.$

Ref.~\cite{Boettcher05d} provided a description of $y_{d}$ as a
continuous function of dimension $d$ using a fit to the data obtained
in Ref.~\cite{Boettcher04b} for $d=2,3,\ldots,6.$ That fit became
credible in that it reproduced the exactly known result in $d=1,$
$y_{1}=-1,$ to within less than 1\%. Hence, it validated the values
for $y_{d}$ found in Refs.~\cite{Boettcher04b,Boettcher04c} and
produced a number of predictions such as that $d_{l}=5/2$ may be the
\emph{lower critical dimension} (the dimension in which
$y_{d}=0)$ for Ising spin glasses, in accordance with an earlier
calculation invoking replica symmetry breaking~\cite{Franz94}.

In a quest for understanding universality in spin glasses, there has
been considerable interest recently in the behavior of $y_{d}$ even
for $d<d_{l},$ where any spin glass ordering is unstable. Presumably,
for divergent energy scales~\cite{Bouchaud03} in Eq.~(\ref{yeq}),
i. e. for $y_{d}>0,$ universality holds and low-temperature properties
of the system are independent of the details of the bond distribution,
as long as it possesses a zero mean and unit variance. In contrast,
below the lower critical dimension significant differences have been
found between classes of bond
distributions~\cite{Hartmann01,Bouchaud03,Amoruso03}.  Especially, for
all $d<d_{l},$ a discrete bond distribution $(J=\pm1)$ leads to
trivial scaling in Eq.~(\ref{yeq}), as was found numerically for
$d=2$~\cite{Hartmann01} and is exactly known for $d=1.$ (In a linear
chain of $L$ spins, the $T=0$ energy difference $\vert\Delta E\vert$
for reversed boundary conditions is given by the smallest bond-weight,
$\vert J\vert=1,$ independent of $L.)$ Only bond distributions $P(J)$
which are continuous near the origin $P(0)$ obtain non-trivial scaling
as represented by the curve in Ref.~\cite{Boettcher05d}, including the
exact result $y_{1}=-1.$ (Here, the smallest bond-weight $\vert
J\vert$ in the chain approaches zero with $1/L.)$

In this paper  we report on a (rather fortuitous) analytical result
for a zero-dimensional spin glass that further clarifies the behavior
of $y_{d<d_{l}}$ for a continuous bond distribution. In the Migdal-Kadanoff
hierarchical lattice (MK)~\cite{Southern77,Bray84,Bouchaud03,Amoruso03}
we find for a $d=0$ dimensional spin glass that $y_{0}=-1$ exactly.

To our knowledge, aside from $d=1$ (where MK is trivially exact)
and the large-link limit~\cite{Bouchaud03}, this is the only exact
result for MK applied to spin glasses. While not of great practical
relevance, studying physical systems in un-physical dimensions has
proved to be of significant theoretical
relevance~\cite{Wilson72,Bender92}. A zero-dimensional spin glass in
particular has in fact been considered previously in Ref.~\cite{Engel93}.

\begin{figure}
\vskip 3.35in
\includegraphics{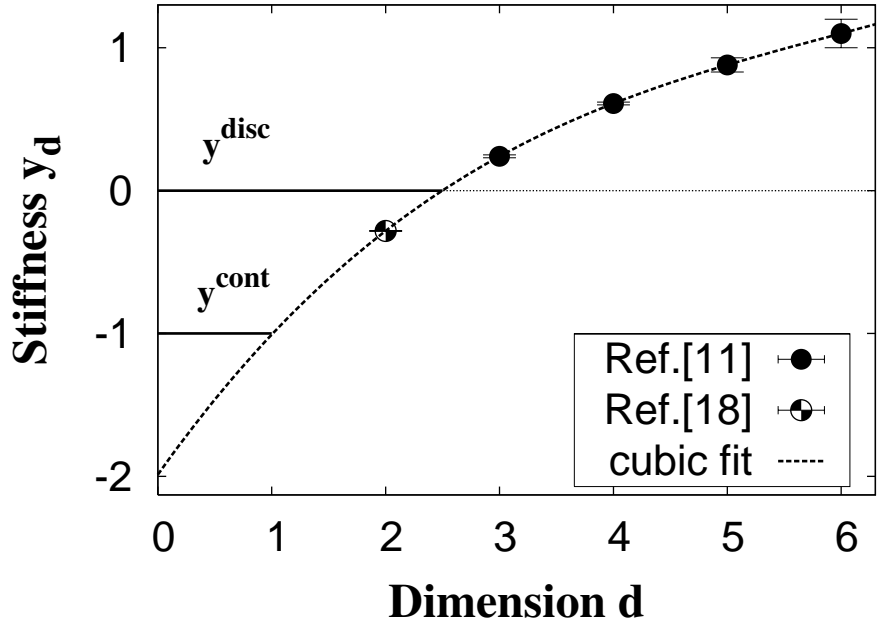}
\caption{Plot of the spin glass stiffness exponent $y_{d}$ as a
function of dimension $d$. Shown are the values for $y_{d}$ on
hyper-cubic lattices from Refs.~\protect\cite{Boettcher04c,Rieger96}
and a cubic fit to that data. For bounded distributions, the exponent
remains locked at $y^{disc}\equiv0$ for $d\leq d_{l}=5/2$ (upper
horizontal line). For continuous distributions, the fit reproduces the
exact result, $y_{1}=-1$ to $0.8\%$ and suggests $y_{0}=-2.$ In
contrast, the result here suggests $y_{0}^{cont}=-1$ and hence,
$y^{cont}\equiv-1$ for all $d\leq1$ (lower horizontal line).}
\label{yd_plot}
\end{figure}

Our result suggests a behavior for $y_{d}$ as depicted in Fig. \ref{yd_plot}:
While for a discrete bond distribution it is $y_{d}^{disc}\equiv0$
for all $d\leq d_{l},$ for a continuous bond distribution, $y_{d}$
first extends smoothly to (non-trivial) negative values through $d_{l}$
towards $d=1,$ beyond which it appears to get fixed at $y_{d}^{cont}\equiv-1$
for all $d\leq1.$ As long as $y_{d}>0$ for $d>d_{l}$ the exponent
is believed to be universal, independent of the bond distribution.

\begin{figure}
\vskip 2.2in
\includegraphics{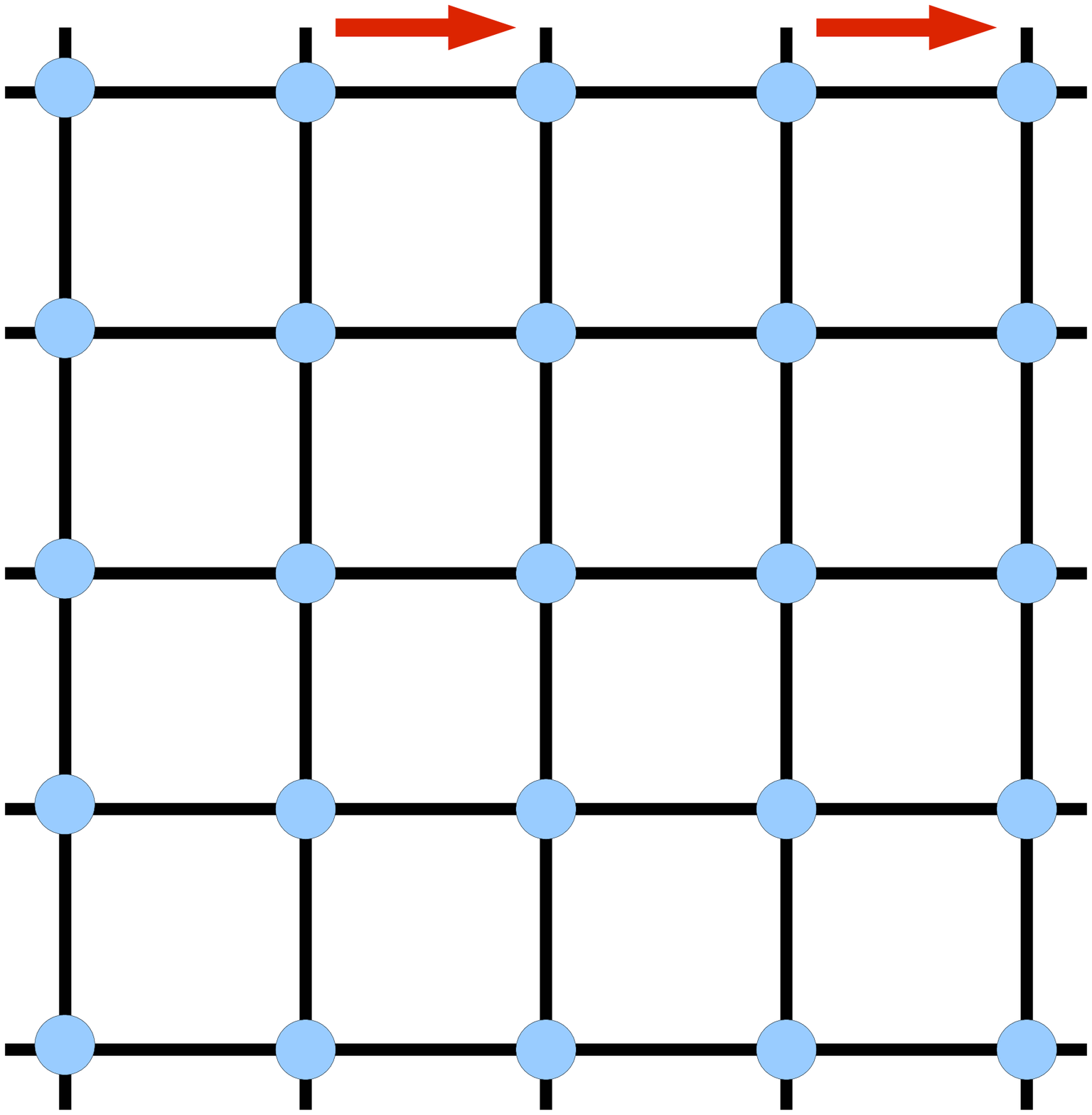}
\includegraphics{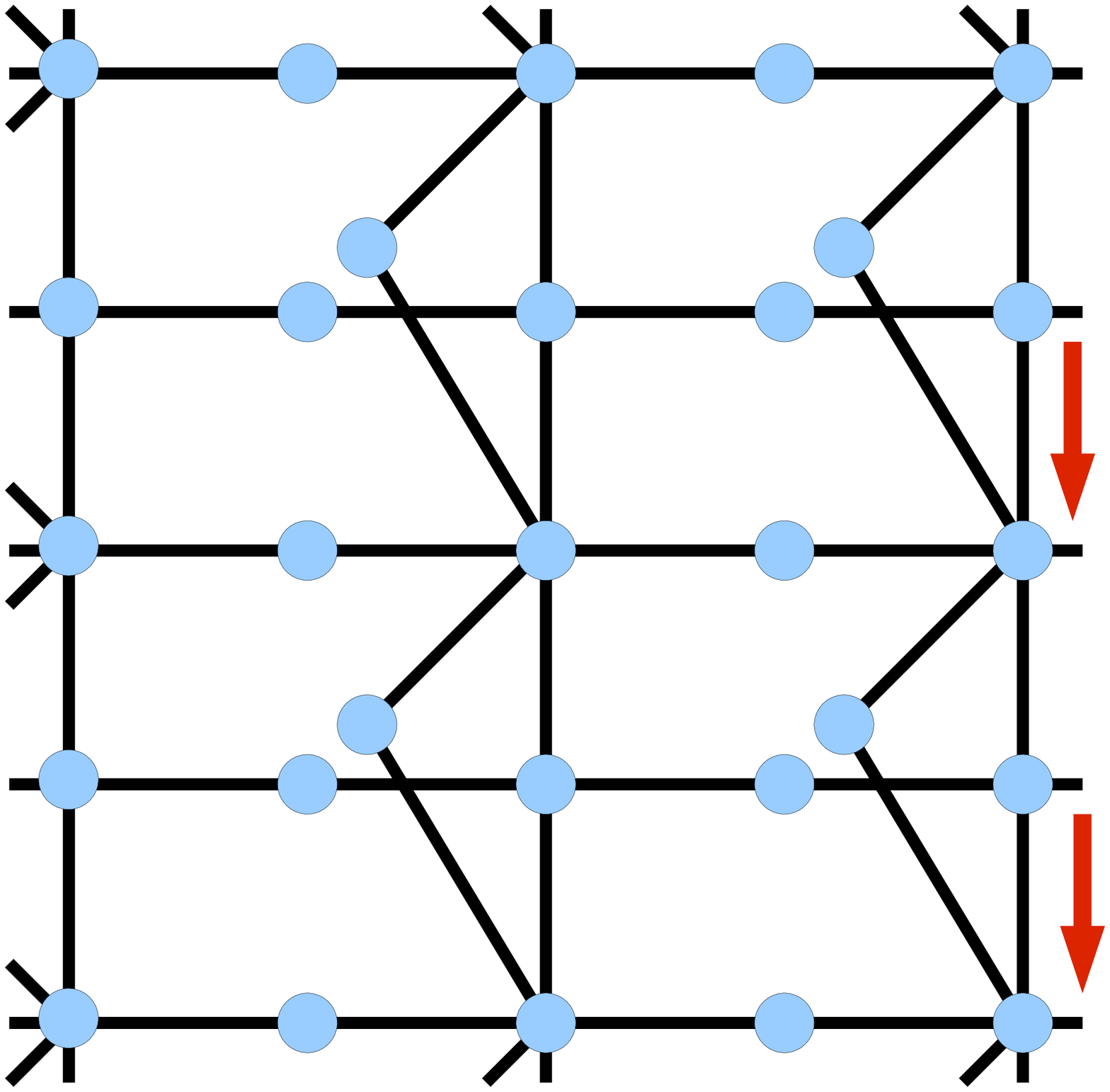}
\includegraphics{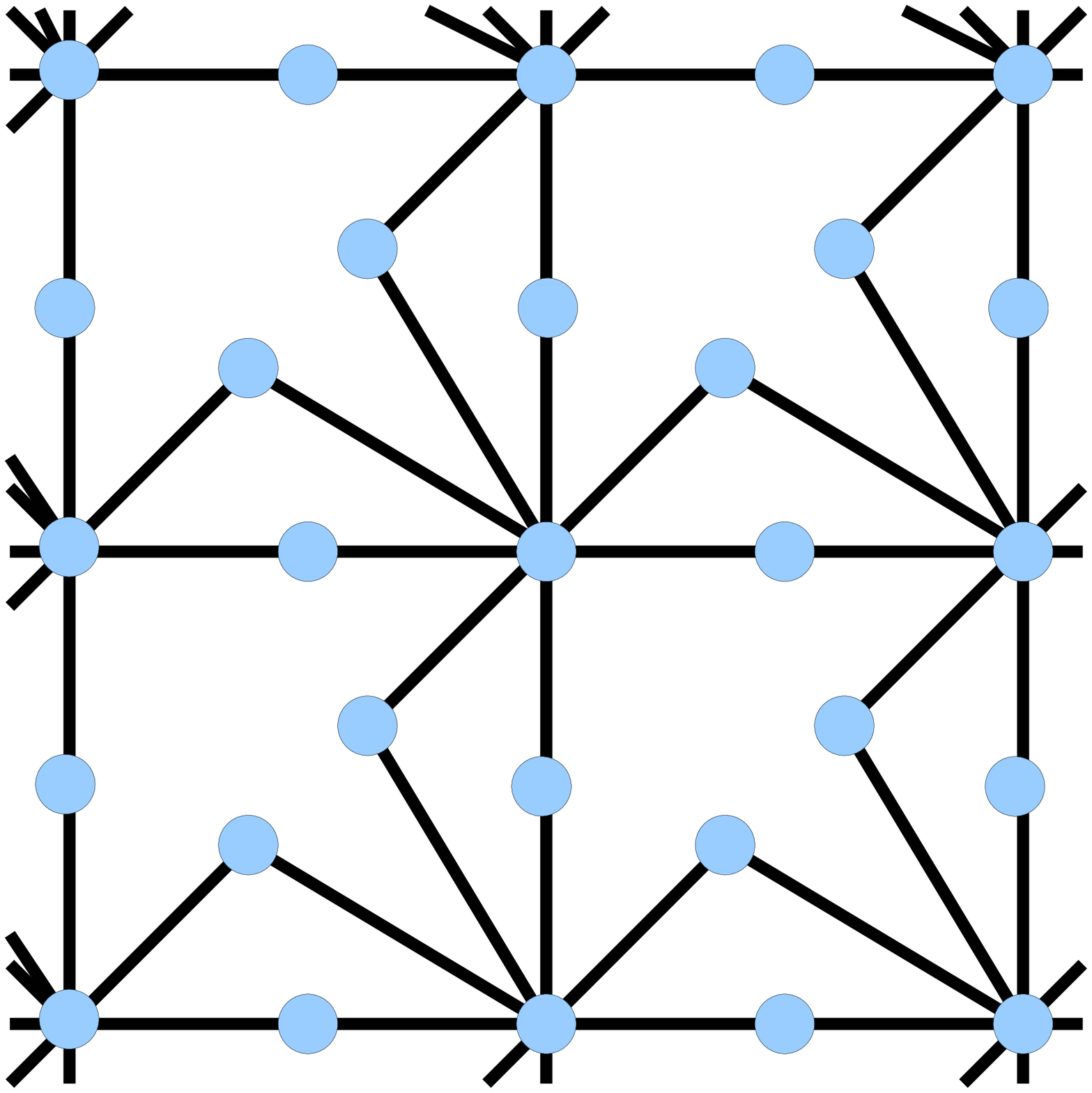}
\caption{Bond-moving scheme in the Migdal-Kadanoff hierarchical
  lattice, here for a square lattice ($d=2$) with $l=2$, i.~e. $b=2$
  in Eq.~(\protect\ref{eq:bl}). Starting from the lattice with unit
  bonds (left), bonds in intervening hyper-planes are projected onto every
  $l$th plane in one direction (middle), then  subsequent directions,
  to re-obtain a similar hyper-cubic lattice of bond-length $l$ (right). The
  renormalized bonds in this case consist of $b=2$ branches, each of a
  series of $l=2$ bonds.}
\label{MKlattice}
\end{figure}

The Migdal-Kadanoff (MK) hierarchical
lattice~\cite{Migdal76,Kadanoff76} provides a real-space
renormalization scheme that approximates especially low-dimensional
spin glasses well and is trivially exact in $d=1.$ These lattices have
a simple recursive, yet geometric, structure and are
well-studied~\cite{Kirkpatrick77,Southern77,Bray84,Bouchaud03}.
Starting from generation $I$ with a single bond, at each subsequent
generation $I+1,$ all bonds from $I$ are replace with a new sub-graph.
This structure of the sub-graph arises from the bond-moving scheme, as
shown in Fig.~\ref{MKlattice}, in $d$
dimensions~\cite{Migdal76,Kadanoff76}: In a hyper-cubic lattice of
unit bond length, at first all $l-1$ intervening hyper-planes of
bonds, transverse to a chosen direction, are projected into every
$l$th hyper-plane, followed by the same step for $l-1$ hyper-planes
being projected onto the $l$th plane in the next direction, and so
on. In the end, one obtains a renormalized hyper-cubic lattice (of
bond length $l)$ in generation $I+1$ with a reformulated $(I+1)$-bond
consisting of a sub-graph of
\begin{equation} 
b=l^{d-1}
\label{eq:bl}
\end{equation} 
parallel branches of a series of $l$ $I$-bonds each. We can rewrite
Eq.~(\ref{eq:bl}) as
\begin{equation} 
d=1+\frac{\ln(b)}{\ln(l)},
\label{eq:deq}
\end{equation} 
anticipating analytic continuation in $l$ and $b$ to obtain results in
arbitrary dimensions $d.$

Instead of solving the problem on the hyper-cubic lattice, we merely
need to consider the recursive scheme of obtaining the
bond-distribution in generation $I+1$ from  sub-graphs of bonds
from generation $I$. Numerically, this is done efficiently at any
temperature $T$ to yield a stationary bond distribution for
$I\to\infty,$ i. e. the thermodynamic limit $L=l^{I}\to\infty.$

Here, we are only concerned with $T=0,$ which simplifies the calculation
drastically to the point that analytical results can be obtained.
For instance, a series of $l$ bonds can always be replace by the bond
of smallest absolute weight. Thus, if these bonds are drawn from a
distribution $P_{I}(J),$ then the distribution $Q_{I}^{(l)}(K)$
of the effective bond $K$ replacing the series can be obtained from
\begin{eqnarray}
Q_{I}^{(l)}(K) & = & {\cal P}\left\{ K=sign\left(J_{1}\times J_{2}\times\ldots\times J_{l}\right)min\left(\left|J_{1}\right|,\ldots,\left|J_{l}\right|\right)\right\} ,\nonumber \\
 & \propto & \int_{-\infty}^{\infty}dJ_{l}P_{I}\left(J_{l}\right)\int_{-\infty}^{\infty}dJ_{l-1}P_{I}\left(J_{l-1}\right)\Theta\left(\left|J_{l}\right|-\left|J_{l-1}\right|\right)\ldots
 \label{eq:Qeq}\\
 &  & \ldots\int_{-\infty}^{\infty}dJ_{1}P_{I}\left(J_{1}\right)\Theta\left(\left|J_{2}\right|-\left|J_{1}\right|\right)\delta\left(\left|J_{1}\right|-\left|K\right|\right),\nonumber 
 \end{eqnarray}
where $\Theta(x)$ refers to the unit step-function and $\delta(x)=\Theta'(x)$
is the Dirac delta-function. In writing Eq.~(\ref{eq:Qeq}), we have
already exploited the symmetry of the integrand under relabeling $J_{1},\ldots,J_{l},$
and we have dropped the obvious norm of $Q^{(l)}(K).$

Similarly, having $b$ bonds drawn from a distribution $Q^{(l)}(K)$
in parallel leads to the distribution $P_{I+1}(J)$ of the reformulated
bond at generation $I+1,$
\begin{eqnarray}
P_{I+1}(J) & = & \int_{-\infty}^{\infty}\prod_{i=1}^{b}\left[dK_{i}\, Q_{I}^{(l)}\left(K_{i}\right)\right]\,\delta\left(J-\sum_{i=1}^{b}K_{i}\right),\nonumber \\
 & = & \int_{-\infty}^{\infty}\frac{d\omega}{2\pi}\, e^{i\omega J}\left[\int_{-\infty}^{\infty}dK\, Q_{I}^{(l)}(K)\, e^{-i\omega K}\right]^{b},
 \label{eq:Peq}
 \end{eqnarray}
where we have used the integral representation $\delta(x)=\int_{-\infty}^{\infty}d\omega\, e^{i\omega x}/(2\pi).$

For general $P_{I},$ Eqs.~(\ref{eq:Qeq}-\ref{eq:Peq}) both are
quite complicated. This is particularly true, since we would be interested
in finding the limiting shape of $P_{I}(J)$ after infinitely many
iterations, $I\to\infty.$ In that limit, we expect the effective coupling
between domains a distance $L=l^{I}$ apart to scale as~\cite{Bray84}
\begin{eqnarray}
{\cal J}_{eff}(L) & \sim & l^{y}\,{\cal J}_{eff}(L/l).
\label{eq:Jeq}
\end{eqnarray}
Each iteration the characteristic {}``width'' ${\cal J}_{eff}(L)\equiv\left[\left\langle J^{2n}\right\rangle_{I}\right]^{1/2n}\sim\left\langle \left|\Delta E\right|\right\rangle _{L},$ of $P_{I}(J)$
increases (or decreases) by a factor of $l^{y},$ so
\begin{eqnarray}
P_{I+1}(J)\sim\frac{1}{l^y}P_I\left(\frac{J}{l^y}\right),\qquad(I\to\infty).
\label{eq:Prob}
\end{eqnarray}
Clearly, if the width grew smaller, $l^{y}<1,$ the behavior of $P_{I}(J)$
near $J=0$ would become increasingly relevant, explaining the non-universal
behavior there~\cite{Bouchaud03}.

We can now report on a solution for Eqs.~(\ref{eq:Qeq}-\ref{eq:Peq})
for \emph{all} values of $I$ for a specific set of initial
distributions and choice of $l$ and $b,$ appropriately analytically
continued. With those choices, only one iteration of
Eqs.~(\ref{eq:Qeq}-\ref{eq:Peq}) is necessary, since the distribution
remains \emph{shape invariant}, i.~e. Eq.~(\ref{eq:Prob}) becomes an
equality.  This set of distributions is continuous and finite near the
origin and thus should represent the universality class containing
Gaussian bonds, for instance. Starting more generally with an initial
distribution
\begin{eqnarray}
P_{0}(J) & =\frac{q+1}{2J_{0}} & \left(1-\frac{\left|J\right|}{J_{0}}\right)^{q}\Theta\left(1-\frac{\left|J\right|}{J_{0}}\right),\qquad(q>-1),
\label{eq:InitialP0}
\end{eqnarray}
where $J_{0}>0$ sets the energy scale for this distribution, it is
easy to show that
\begin{eqnarray}
Q_{0}^{(l)}(K) & = & \frac{l\left(q+1\right)}{2J_{0}}\left(1-\frac{\left|K\right|}{J_{0}}\right)^{lq+l-1}\Theta\left(1-\frac{\left|K\right|}{J_{0}}\right),
\label{eq:InitialQ}
\end{eqnarray}
by recursion of Eq.~(\ref{eq:Qeq}). Note, that Eq.~(\ref{eq:InitialQ})
readily continues to any real value of $l>0.$

The evaluation of Eq.~(\ref{eq:Peq}) is somewhat more complex. To
facilitate the subsequent analysis, it is best to consider the moment
generating function for $P_{I}(J),$
\begin{eqnarray}
\phi_{I+1}\left(\alpha\right) & = & \left\langle e^{-i\alpha J}\right\rangle _{I+1},\nonumber \\
 & = & \int_{-\infty}^{\infty}dJ\, e^{-i\alpha J}\int_{-\infty}^{\infty}\frac{d\omega}{2\pi}\, e^{i\omega J}\left[\int_{-\infty}^{\infty}dK\, Q_{I}^{(l)}(K)\, e^{-i\omega K}\right]^{b},\\
 & = & \left[\int_{-\infty}^{\infty}dK\, Q_{I}^{(l)}(K)\, e^{-i\alpha K}\right]^{b},\nonumber 
\label{eq:GenFunc}
\end{eqnarray}
where the integral over $J$ merely represents a $\delta-$function.
Using Eq.~(\ref{eq:InitialQ}), we find
\begin{eqnarray}
\phi_{1}\left(\alpha\right) & = & \left[\int_{-\infty}^{\infty}dK\, Q_{0}^{(l)}(K)\, e^{-i\alpha K}\right]^{b},\nonumber \\
 & = & \left[l\left(q+1\right)\int_{0}^{1}dx\,\left(1-x\right)^{lq+l-1}\cos\left(\alpha J_{0}x\right)\right]^{b}.
\label{eq:Phi1}
\end{eqnarray}
Correspondingly, we find for the generating function of the initial
bond distribution in Eq.~(\ref{eq:InitialP0}),
\begin{eqnarray}
\phi_{0}(\alpha) =  \left\langle e^{-i\alpha J}\right\rangle _{0},
 =  \left(q+1\right)\int_{0}^{1}dx\,\left(1-x\right)^{q}\cos\left(\alpha J_{0}x\right).
\label{eq:Phi0}
\end{eqnarray}

Despite the obvious similarities between
Eqs.~(\ref{eq:Phi1}-\ref{eq:Phi0}), finding a set of parameters that
make $\phi_{0}$ and $\phi_{1}$ similar are hard to find because of the
exponent $b$ in Eq.~(\ref{eq:Phi1}).  Even for $b=1$ and $l\not=1,$
i. e. $d=1$ according to Eq.~(\ref{eq:deq}), we need to iterate for
$I\to\infty$ to find the width to decay to zero with $1/L,$
i. e. $y_{1}=-1.$ It is merely a lucky circumstance that there are
two solution with the required properties, each independently
yielding the same result.

First, for $l=2$ we have to continue the {}``branching number'' to
$b=1/2$ to obtain $d=0$ in Eq.~(\ref{eq:deq}). Then, for the
rectangular function for $P_0(J)$, i.~e. $q=0$, we find
\begin{eqnarray} 
\phi_{0}(\alpha)=\frac{\sin\left(\alpha
J_{0}\right)}{\alpha J_{0}}, & \qquad &
\phi_{1}\left(\alpha\right)=\left[\frac{4\sin^{2}\left(\frac{\alpha
J_{0}}{2}\right)}{\left(\alpha J_{0}\right)^{2}}\right]^{\frac{1}{2}},
\label{eq:firstcase}
\end{eqnarray} 
both of which are invariant in an sufficiently large open interval
around $\alpha=0,$ required to generate any moment. We identifying in
Eq.~(\ref{eq:Jeq}) ${\cal J}_{eff}(L=l^{0})=J_{0}$ and ${\cal
J}_{eff}(L=l^{1})=J_{0}/2=l^{y}{\cal J}_{eff}(L=l^{0}),$ hence,
$y_{0}=-1.$

Second, for $b=2$ we have to continue the ``series number'' to $l=1/2$
to obtain $d=0$ in Eq.~(\ref{eq:deq}). Then, for the triangular
function for $P_0(J)$, i.~e. $q=1$, we find
\begin{eqnarray}
\phi_{0}\left(\alpha\right)=
\frac{4\sin^{2}\left(\frac{\alpha J_{0}}{2}\right)}{\left(\alpha J_{0}\right)^{2}}, & \qquad &
\phi_{1}(\alpha)=\frac{\sin^{2}\left(\alpha J_{0}\right)}{\left(\alpha
    J_{0}\right)^{2}}.
\label{eq:secondcase}
\end{eqnarray}
In this case, we identify in Eq.~(\ref{eq:Jeq}) again ${\cal
J}_{eff}(L=l^{0})=J_{0},$ but ${\cal J}_{eff}(L=l^{1})=2J_{0},$ which
now results again in $y_{0}=-1$ because length-scales are actually
shrinking, $l=1/2.$

It is clear that nothing will change on this result under further
iteration, $I\to I+1.$ Assuming that there is a unique solution for
Eq.~(\ref{eq:Prob}), one would expect that any choice for $P_0(J)$
that is continuous at $J=0$ should converge to a rectangular
(triangular) function for $l=1/b=2$ ($=\frac{1}{2}$). We have not been able
to extract any further result of this nature.

Hence, we are lead to believe that $y_{0}=-1.$ Of course, such a
result has to be considered with care. It is a considerable weakness
of MK that its results for a given dimension $d$ according to
Eq.~(\ref{eq:deq}) are not generally unique~\cite{Boettcher05d}.
Here, at least, we found two different combinations of $l$ and $b,$
both giving $d=0$ and yielding an identical stiffness exponent. This
may indicate a unique result for any combination $l=1/b.$ Furthermore,
it is not clear that the MK result for $d=0$ should necessarily correspond
{\it quantitatively} to an Edwards-Anderson spin glass on a zero-dimensional
lattice. This is only known to be true for $d=1$ (i. e. $b=1)$ and certainly
wrong for any $d>1,$ albeit with slowly increasing error.

This work has been supported by grant 0312510 from the Division of
Materials Research at the National Science Foundation and by the Emory
University Research Council. 

\comment{
These commands specify that your bibliography should be created from entries in the file bibdata.bib. There are several bibliography styles that can be used with the eethesis document class; the style theunsrt.bst is the style used in the examples in this manual. The entries in this style are patterned after those in IEEE Transactions on Automatic Control. It lists the sources in the order they were cited. There is also a generic ieeetr.bst, which formats sources similar to many IEEE publications. If neither of these styles is suitable for your department, you might consider acm.bst or siam.bst which format your bibliography in the style of ACM and SIAM publications. Check the /usr/local/teTeX/local.texmf/bibtex/bst directory on the machine that you use to see if there are any other .bst files you can use.} 

\bibliographystyle{apsrev}
\bibliography{dzero}

\end{document}